\documentclass[reprint,aps,superscriptaddress,footinbib,prl]{revtex4-2}
\usepackage{amsmath}
\usepackage{physics}
\usepackage{graphicx}
\usepackage{amssymb}
\usepackage{amsthm}
\usepackage{bm}
\usepackage{txfonts}
\usepackage{pstricks}
\usepackage[colorlinks=true,allcolors=blue]{hyperref} 
\usepackage[capitalise]{cleveref}
\allowdisplaybreaks

\newcommand{\vk}{{\bm{k}}}
\newcommand{\vq}{{\bm{q}}}
\newcommand{\vp}{{\bm{p}}}
\newcommand{\vm}{{\bm{m}}}

\definecolor{mygreen}{rgb}{0.15, 0.6, 0.15}
\definecolor{mygrey}{rgb}{0.5, 0.5, 0.5}

\begin{document}
\title{Highly accurate electronic structure of metallic solids from coupled-cluster theory with nonperturbative triple excitations}
\author{Verena A. Neufeld}
\affiliation{Department of Chemistry, Columbia University, New York, New York 10027, USA}
\author{Timothy C. Berkelbach}
\affiliation{Department of Chemistry, Columbia University, New York, New York 10027, USA}
\date{\today}  

\begin{abstract}
Coupled-cluster theory with single, double, and perturbative triple excitations
(CCSD(T))---often considered the ``gold standard'' of main-group quantum
chemistry---is inapplicable to three-dimensional metals due to an infrared
divergence, preventing its application to many important problems in materials
science.
We study the full, nonperturbative inclusion of triple excitations (CCSDT) and
propose a new, iterative method, which we call ring-CCSDT, that resums the
essential triple excitations with the same $N^7$ run-time scaling as CCSD(T).
CCSDT and ring-CCSDT are used to calculate the correlation energy of the
uniform electron gas at metallic densities and the structural properties of
solid lithium.  Inclusion of connected triple excitations is shown to be
essential to achieving high accuracy.  We also investigate semiempirical CC
methods based on spin-component scaling and the distinguishable cluster
approximation and find that they enhance the accuracy of their parent
\textit{ab initio} methods.
\end{abstract}

\maketitle

\textit{Introduction.}
Accurately predicting energetic properties of metallic solids is crucial in
computational materials science, with applications in heterogeneous
catalysis, electrochemistry, and battery 
science~\cite{norskov_density_2011,calle-vallejo_first-principles_2012,he_density_2019}.
Coupled-cluster theory with single and double excitations
(CCSD)~\cite{shavitt_many-body_2009,bartlett_coupled-cluster_2007} has recently
been shown to provide reasonable energies for the uniform electron gas
(UEG)~\cite{shepherd_communication_2016,mihm_power_2021,callahan_dynamical_2021}
and for atomistic metallic solids, such as lithium and
aluminum~\cite{stoll_incremental_2009,mihm_shortcut_2021,neufeld_ground-state_2022,weiler_machine_2022},
but it does not reliably outperform density functional theory (DFT), which is
significantly cheaper---some inclusion of connected triple excitations is clearly required.
For non-metallic main-group solids, CCSD with perturbative triple excitations
(CCSD(T))~\cite{raghavachari_fifth-order_1989} is highly accurate
for bulk properties~\cite{schwerdtfeger_convergence_2010,booth_towards_2013,gruneis_coupled_2015,gruber_applying_2018,gruber_ab_2018} and surface
chemistry~\cite{tsatsoulis_comparison_2017,gruber_ab_2018,tsatsoulis_reaction_2018,brandenburg_physisorption_2019,lau_regional_2021},
mirroring its performance on molecules, where it commonly yields ``chemical accuracy'' of
about 1~kcal/mol~\cite{bartlett_coupled-cluster_2007}.
However, CCSD(T) is not expected to be applicable to three-dimensional metals: 
an approximate evaluation of the CCSD(T) energy of the UEG was shown to diverge in 
the thermodynamic limit~\cite{shepherd_many-body_2013}, similar to the textbook result of second-order
perturbation theory~\cite{macke_uber_1950,gell-mann_correlation_1957}.

Here, we investigate the accuracy of CC theory with nonperturbative triple
excitations (CCSDT)
to determine whether such a theory provides the desired accuracy for metals.
Because the high cost of CCSDT limits its routine application, we also design and test lower cost
alternatives.
Below, we first review diagrammatic results on the ground-state
energy of the UEG, including its high-density expansion, divergences and necessary
resummations, and connections with coupled-cluster theory including double and triple excitations.
An analysis of the (T) correction for the UEG motivates a new theory, which nonperturbatively
retains the triple excitations necessary to preclude a divergence and which has the same
$N^7$
computational scaling as CCSD(T).
We assess the performance of these methods with applications to the UEG at metallic densities
and to solid lithium.
Furthermore, we test several empirical modifications, including the
distinguishable cluster (DC)
approximation~\cite{kats_communication_2013,kats_distinguishable_2019,rishi_can_2019}
and spin-component-scaled (SCS) CC
theory~\cite{grimme_improved_2003,takatani_improvement_2008,kats_improving_2018},
which were designed to approximate the effect of higher excitations without
increasing the computational cost.

\textit{Diagrammatic results on the uniform electron gas.}
The UEG, a model of interacting electrons in a uniform positive background, 
has been a famous testing ground for new developments in
nonperturbative many-body quantum field theory.  Specifically, the total energy
of the UEG with electron density $n$ has been evaluated to
leading orders in the Wigner-Seitz radius 
$r_s = (3/(4\pi n))^{1/3}$~\cite{gell-mann_correlation_1957,carr_ground-state_1964,Endo1999},
in the absence and presence of a spin polarization; in this work, we focus on the upolarized
case.
The kinetic energy and Hartree-Fock exchange energy produce terms
of $O(r_s^{-2})$ and $O(r_s^{-1})$, respectively, and
the remaining terms define the correlation energy.

\begin{figure}[t]
	\includegraphics[width=3.25in]{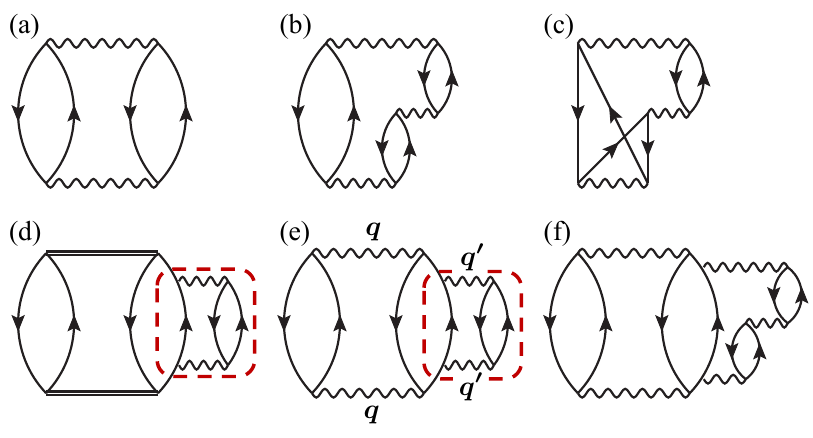}
	\caption{
            Goldstone diagrams discussed in the text, which are included at various orders in perturbation
            theory and various flavors of CC theory. The dashed red box in (d) and (e) highlights the problematic
            feature responsible for the divergence of the CCSD(T) correlation energy.
        }
	\label{fig:diagrams}
\end{figure}

From dimensionality arguments, it is expected that second-order perturbation theory
contributes all terms of $O(r_s^0)$, which is correct for the second-order
exchange energy~\cite{gell-mann_correlation_1957,onsager_integrals_1966}.
The second-order direct (ring) term, whose diagram is shown in
Fig.~\ref{fig:diagrams}(a), contributes a correlation energy
$E_{24} \propto r_s^0 \int_0^\infty dq f(q)/q^2$, where
\begin{equation}
f(q) = \int_{|\vk+\vq|>1}d^3k \int_{|\vp+\vq|>1}d^3p
    \frac{\theta(1-k)\theta(1-p)}{q^2 + (\vk+\vp)\cdot\vq}
\end{equation}
and all dimensionless momenta $\vk, \vp, \vq$ are normalized to the Fermi momentum;
we use the notation $E_{m,2n}$ from Refs.~\onlinecite{carr_ground-state_1964,Endo1999},
where $m$ is the order in perturbation theory and $n$ is the number of interactions with
the same momentum transfer.
It can be shown that $f(q) \propto q$ in the limit $q\rightarrow 0$, and thus
the second-order direct term famously diverges logarithmically.  All higher order terms
with the same ring structure ($n$ rings at order $n$ in
perturbation theory), such as the one shown in Fig.~\ref{fig:diagrams}(b) (i.e., $E_{36}$) 
exhibit the strongest divergences at each order, and their resummation to infinite order
defines the random-phase approximation
(RPA)~\cite{macke_uber_1950,bohm_collective_1951,pines_collective_1952,bohm_collective_1953,gell-mann_correlation_1957,carr_ground-state_1964,Endo1999},
$\varepsilon' = E_{24} + E_{36} + E_{48} + \cdots$.
The RPA provides a correlation energy that is correct to $O(\ln r_s)$ and is
therefore exact in the high-density $r_s\rightarrow 0$ limit (aside from a constant); 
the appearance of terms $O(\ln r_s)$ in the density expansion signals the non-analyticity of the
correlation energy.  As is well-known, the CCSD energy contains all terms
included in the
RPA~\cite{freeman_coupled-cluster_1977,bishop_electron_1978,scuseria_ground_2008},
providing a strong theoretical argument for the application of CC theories to
metallic solids---a research agenda started more than 40 years
ago~\cite{freeman_coupled-cluster_1977,bishop_electron_1978,bishop_electron_1982,emrich_electron_1984}.


Third-order perturbation theory produces 
convergent terms that are $O(r_s)$ (i.e., $E_{32}$),
strongly divergent terms with three rings that are included in the RPA [i.e., Fig.~\ref{fig:diagrams}(b)
or $E_{36}$],
and more weakly divergent terms whose diagrams have only one ring,
such as that shown in Fig.~\ref{fig:diagrams}(c), which define $E_{34}$.
These latter terms have to be resummed with higher-order divergent contributions
that have analogous structure ($n-2$ rings at order $n$ in perturbation theory), 
$\varepsilon'' = E_{34} + E_{48} + \cdots$,
which can be evaluated to identify a correlation energy that is exact
to $O(r_s, r_s\ln r_s)$~\cite{dubois_electron_1959,carr_ground-state_1964,Endo1999}.
Remarkably, all of these terms are included in the CCSD correlation energy.
Although it has long been appreciated that CCSD resums the most divergent
terms that define the RPA correlation energy $\varepsilon'$~\cite{freeman_coupled-cluster_1977,bishop_electron_1978,scuseria_ground_2008}, to the
best of our knowledge, it has not been noted that it also resums these
next most divergent terms that define $\varepsilon''$.
Therefore, CCSD is exact for the energy of the UEG to $O(r_s, r_s\ln r_s)$,
which is one order higher than the RPA, in addition to recovering the correct
constant term due to second-order exchange.

As expected, the CCSD energy is missing terms from fourth order in perturbation
theory, including those that yield finite values of $O(r_s^2)$ or that diverge
weakly and must be resummed with higher-order terms. CCSDT produces an energy that
is exact to fourth order in perturbation theory and includes resummations necessary to
eliminate fourth-order divergences, thus providing
a potentially powerful theory of the energy of metals.  However, CCSDT has a high
computational cost that scales as $N^8$, which precludes routine application to
atomistic materials.
Nonetheless, below we exploit the simplicity of the UEG and carefully
designed composite corrections to provide the first estimates of the
performance of CCSDT for the UEG in the thermodynamic limit and for solid lithium. 

The intermediate theory CCSD(T), with a reduced $N^7$
scaling, is very accurate for many molecules and insulating solids. However,
CCSD(T) yields a divergent energy for metals, which was demonstrated
numerically using an approximate form in
Ref.~\onlinecite{shepherd_many-body_2013}.  Here, we provide a diagrammatic analysis of
the same behavior to shed more light on the failures of CCSD(T).
Neglecting single excitations, which vanish for the UEG by symmetry, the energy
correction in CCSD(T) is shown by the diagram in Fig.~\ref{fig:diagrams}(d)
(plus permutations due to exchange),
where the double line indicates a converged CCSD $T_2$ amplitude.  To lowest order,
the (T) correction is that of bare fourth-order perturbation theory, shown in
Fig.~\ref{fig:diagrams}(e), whose analysis elucidates the (T) divergence.
Considering only the contribution without exchange, the problematic process
has four interactions with two pairs of identical momenta exchanged, $\vq$ and $\vq'$,
i.e., the correlation energy is
$E_\mathrm{c} \propto r_s^2 \int d^3 q \int d^3 q' f(\vq,\vq')/(q^4 q'^4)$,
where
\begin{widetext}
\begin{equation}
\label{eq:e4}
f(\vq,\vq') = \int_{|\vk+\vq|>1}d^3k \int_{|\vm+\vq'|>1}d^3m
	\int_{\substack{|\vp+\vq|>1\\|\vp-\vq'|<1}}d^3p 
	\frac{\theta(1-k)\theta(1-p)\theta(1-m)}
        {[q^2 + (\vk+\vp)\cdot\vq]^2 [q^2 + (\vk+\vp)\cdot\vq + (\vm+\vp)\cdot\vq']}.
\end{equation}
\end{widetext}
As usual, the correlation energy integral diverges due to the behavior of the 
integrand near $q,q'=0$.
Letting $q_c$ be an infrared cutoff on both momentum integrals, the integrated
result can be checked to diverge
as $O(q_c^{-2} \ln q_c)$, demanding resummation with higher-order terms.

By replacing the outer Coulomb interactions by CCSD $T_2$ amplitudes as in
Fig.~\ref{fig:diagrams}(d), CCSD(T) regularizes the integral over $\vq$, but
not $\vq'$.  This single ring diagram self-energy insertion, highlighted with a
red box in Figs.~\ref{fig:diagrams}(d) and (e), is responsible for the
divergence of the CCSD(T) energy for metals. By analytically performing this
regularization, the CCSD(T) energy can be shown to diverge as $O(\ln q_c)$,
which is naturally weaker than that of bare fourth-order perturbation theory,
but still useless for quantitative calculations. 
This rate of divergence is exactly the same as that of second-order
perturbation theory, which we exploit in the Supplemental
Material~\footnote{See Supplemental Material at [URL will be inserted by
publisher] for technical details of all calculations and a discussion
of finite-size errors in metals, including numerical demonstration of
convergent and divergent behaviors of the considered theories.} 
to numerically confirm the divergence of CCSD(T), along the lines of
other works~\cite{shepherd_range-separated_2014,shepherd_coupled_2014}.

Importantly, this analysis also identifies the minimal physics necessary to
regularize the CCSD(T) approximation for metals, which is an infinite-order
RPA-style resummation of ring diagrams in the self-energy insertion (like in
the GW approximation~\cite{hedin_new_1965}), as shown in
Fig.~\ref{fig:diagrams}(f).  This can be achieved approximately by removing
many of the terms from the CCSDT equations, analogous to the equivalence
between (direct) ring-CCD and the RPA.  This method, which we call ring-CCSDT,
is implemented as follows.  The singles and doubles amplitude equations are
exactly as in CCSDT.  The triples amplitude equation is the same as in the
CCSDT-1
approximation~\cite{lee_study_1984,lee_coupled_1984,urban_towards_1985,noga_towards_1987,shavitt_many-body_2009},
but is supplemented with direct ring diagrams,
$\epsilon_{ijk}^{abc} t_{ijk}^{abc}  = R_{\mathrm{CCSDT-1}} +  R_{\mathrm{dr}}$,
\begin{subequations}
\begin{align}
\begin{split}
&R_{\mathrm{CCSDT-1}} = 
\hat{P}(k/ij|a/bc) \langle bc || dk \rangle t_{ij}^{ad} 
- \hat{P}(i/jk|c/ab) \langle lc||jk \rangle t_{il}^{ab} \\
     &\hspace{1em}+ 
\hat{P}(c/ab) (f_{cd} - \epsilon_c \delta_{cd}) t_{ijk}^{abd} 
- \hat{P}(k/ij) (f_{lk} - \epsilon_k \delta_{lk}) t_{ijl}^{abc} 
\end{split} \\
\begin{split}
&R_{\mathrm{dr}} = \hat{P}(i/jk|a/bc) \bra{al}\ket{id} t_{ljk}^{dbc} 
+ \hat{P}(i/jk|abc) \bra{lb}\ket{de} t_{il}^{ad} t_{jk}^{ec} \\
    &\hspace{1em} 
- \hat{P}(ijk|a/bc) \bra{lm}\ket{dj} t_{il}^{ad}t_{mk}^{bc}
+ \hat{P}(i/jk|a/bc) \bra{lm} \ket{de} t_{il}^{ad} t_{mjk}^{ebc}
\end{split}
\end{align}
\end{subequations}
where
$\hat{P}(k/ij|a/bc) = [1 - \hat{P}(ik) - \hat{P}(jk)][1 - \hat{P}(ab) - \hat{P}(ac)]$
and $\hat{P}(ij)$ generates the permutation of $i$ and $j$
(as usual, $\varepsilon_{ijk}^{abc}$ is an orbital energy denominator, 
$f_{pq}$ is a Fock matrix element,
$i,j,k,l,m$ indicate occupied spin orbitals, $a,b,c,d,e$ unoccupied
spin orbitals, Coulomb integrals are in $\langle 12|12\rangle$ notation,
the double bar indicates antisymmetrized integrals, and summation over
repeated indices $l,m,d,e$ is implied).

Unfortunately, despite its iterative nature,
the CCSDT-1 approximation (without the ring diagrams)
is a divergent theory of metals, like CCSD(T),
because of the isolated ring diagram highlighted in Figs.~\ref{fig:diagrams}(d)
and (e).
In the ring-CCSDT approximation,
not all time-orderings of repeated ring diagrams are included: all forward (Tamm-Dancoff)
time-orderings are included, which is sufficient to preclude a divergence~\cite{freeman_coupled-cluster_1977},
and a subset of the non-Tamm-Dancoff time-orderings are included, but not
all those corresponding to the complete RPA; this is very similar to the diagrammatic
content of the coupled-cluster Green's function~\cite{lange_relation_2018,tolle_exact_2022}.
To include all time-orderings that define RPA screening would require inclusion
of connected quadruple excitations.

The first and last terms of $R_\mathrm{dr}$ exhibit $N^8$ computational scaling, like the
parent CCSDT method. However, the use of direct (non-antisymmetrized) ring 
diagrams enables a reduction in scaling with the use of density-fitting
(or Cholesky decomposition)
of the Coulomb integrals $\bra{pq}\ket{rs} = \sum_P L_{pr}^{P} L_{qs}^{P}$,
where $P$ is an auxiliary index.
For example, the last term can be constructed as
\begin{equation}
\sum_{lmde} \bra{lm} \ket{de} t_{il}^{ad} t_{mjk}^{ebc} 
= \sum_P \left[\sum_{ld} L_{ld}^P  t_{il}^{ad}\right] \left[ \sum_{me} L_{me}^P t_{mjk}^{ebc}\right].
\end{equation}
With such a compression of the Coulomb integrals, ring-CCSDT is an iterative
$N^7$ method, providing an appealing alternative to the CCSD(T) approximation
that is applicable to metals (although the storage of the $T_3$ amplitudes
is a separate bottleneck).

\begin{figure}[t]
	\includegraphics[width=3.25in]{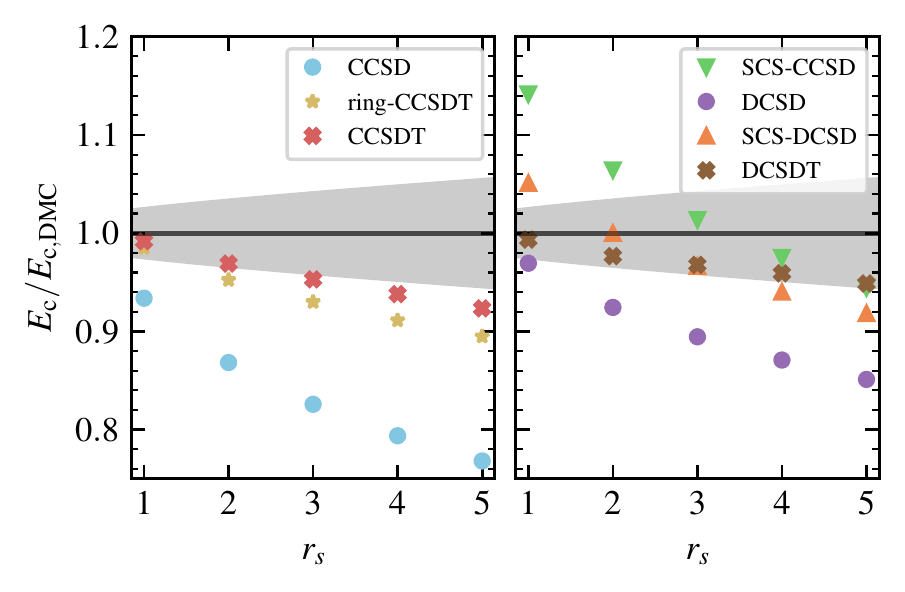}
        \caption{Ratio of the coupled-cluster correlation energy to the diffusion Monte
            Carlo (DMC) correlation
            energy~\cite{ceperley_ground_1980,perdew_self-interaction_1981} for
            the three-dimensional UEG with $r_s = 1$--5, as given by the
            methods indicated in the legend.
            The methods are separated into those that are purely 
            diagrammatic (left) and those that are semi-empirical (right).
            Range of chemical accuracy ($\pm$1~kcal/mol or $\pm$1.6~m$E_h$) is shown with a grey shaded area.
        }
	\label{fig:ueg_rs_frac}
\end{figure}

\textit{Results for the UEG.}
CC approximations are difficult to treat semi-analytically,
even for the UEG.  Therefore, we simulate a UEG of electron density $n$ via a cubic box
of $N$ electrons with volume $V=N/n=(4/3)\pi r_s^3N$ and a plane-wave orbital basis.  
Although several improved CC methods have been previously applied to UEG models
containing a finite number of
electrons~\cite{mcclain_spectral_2016,spencer_developments_2016,neufeld_study_2017,liao_towards_2021}, here
we are concerned with the critical question of their performance in the
thermodynamic limit, which we estimate via basis set corrections and
extrapolations to the thermodynamic limit. 
Specifically, we perform CCSD and DCSD calculations on systems
containing up to $N=1404$ electrons and estimate the complete basis set limit
using calculations on smaller system sizes. These results are then used to
extrapolate to the thermodynamic limit assuming that finite-size errors in the correlation
energy decay asymptotically as $N^{-2/3}$---a functional form that has also
been proposed in recent work~\cite{mihm_how_2023}. 
Our final CCSD correlation energies agree within about 1~m$E_h$ with previous
studies that targeted the thermodynamic
limit~\cite{shepherd_communication_2016,mihm_power_2021}, despite different
technical details, providing a validation of our methods.  
CCSDT, ring-CCSDT, and DCSDT calculations are performed on systems containing up to
$N=156$ electrons, and we calculate the energy difference with respect to DCSD.
The complete basis set limit of this energy difference is estimated based on smaller
values of $N$ and then extrapolated to the thermodynamic limit.
Additional technical details are given in the Supplemental Material~\cite{Note1}.

In Fig.~\ref{fig:ueg_rs_frac}, we present the correlation energy of the UEG at
metallic densities of $r_s=1$--5 from various CC theories 
as a fraction of the numerically exact result, estimated via the
Perdew-Zunger fit~\cite{perdew_self-interaction_1981} to diffusion Monte Carlo
(DMC) results~\cite{ceperley_ground_1980};
a table of all values is given in the Supplemental Material~\cite{Note1}.
The magnitude of the DMC correlation energy ranges from 60~m$E_h$ at $r_s=1$ to 28~m$E_h$
at $r_s=5$. 
As expected based on the density expansion discussed above, the relative
accuracy of diagrammatic methods shown in Fig.~\ref{fig:ueg_rs_frac}(a) (CCSD,
CCSDT, and ring-CCSDT) decreases with increasing $r_s$.
Compared to CCSD, which recovers only about 75--95\% of the DMC correlation energy,
CCSDT performs extremely well and recovers between 99\% (at $r_s=1$) and 92\% (at $r_s=5$),
corresponding to an absolute accuracy of 0.5--2.2~m$E_h$.
The good performance of ring-CCSDT, with errors of 0.9--3.0~m$E_h$, 
shows that the same ring diagram resummation responsible for curing the divergence
of CCSD(T) is also responsible for most of the correlation energy associated with
connected triple excitations.

The semi-empirical CC methods shown in Fig.~\ref{fig:ueg_rs_frac}(b) (SCS-CCSD, DCSD,
SCS-DCSD, and DCSDT) typically 
perform better than their parent diagrammatic method. SCS-CCSD~\cite{takatani_improvement_2008}
improves over CCSD, except at small $r_s$, demonstrating that semi-empirical modifications can spoil valuable
formal properties like the exactness of CC theories in the high-density limit.
DCSD~\cite{kats_communication_2013} is better behaved and roughly halves the error of CCSD over this density range.
SCS-DCSD~\cite{kats_improving_2018} is a further improvement and provides the best overall performance of the
$N^6$ scaling methods. Remarkably, DCSDT~\cite{kats_distinguishable_2019,rishi_can_2019}
yields results of extremely high accuracy,
recovering more than 94\% of the DMC correlation energy at all densities, which
corresponds to an error of less than 1.6~m$E_h$, i.e., under 1~kcal/mol.

\textit{Results on solid lithium.}
Next, we investigate the transferability of the above performance to a real
material.  We study solid lithium, which is a simple metal with a valence
electron density corresponding to $r_s \approx 3.2$.  We use CCSD, DSCD,
ring-CCSDT, CCSDT, and DCSDT to calculate the equilibrium lattice parameter,
bulk modulus, and cohesive energy.  All calculations were performed with a
development branch of
PySCF~\cite{sun_libcint_2015,sun_pyscf_2018,sun_recent_2020}, and all technical
details---such as pseudopotentials, basis sets (up to quadruple-zeta Gaussian
type orbitals), and Brillouin zone samplings (up to 64 $k$-points, plus
extrapolation)---are the same as in our previous
work~\cite{neufeld_ground-state_2022}; in that work, we found that CCSD
predictions had significant room for improvement (at the CCSD level, we find
that our updated finite-size extrapolations cause only small differences from
our previous work, e.g., about 0.8~m$E_h$ in the cohesive energy).
We estimate the ring-CCSDT, CCSDT, and DCSDT energies
using composite corrections, by again considering the differences to DCSD,
based on calculations with small supercells
(containing 8 and 16 Li atoms), frozen core orbitals, and frozen virtual
natural orbitals~\cite{Note1}.

\begin{figure}[t]
	\includegraphics[width=3.25in]{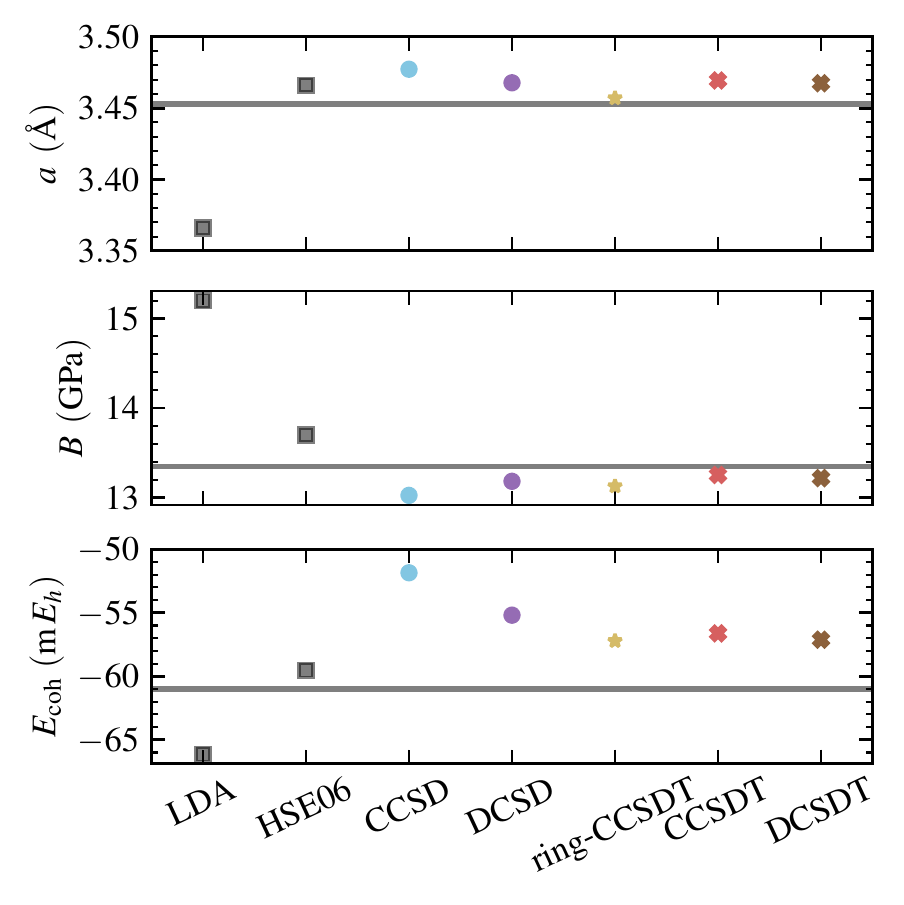}
        \caption{Equilibrium lattice constant $a$, bulk modulus $B$, and
            cohesive energy $E_{\mathrm{coh}}$ for solid lithium.  Results are
            shown at the indicated levels of CC theory and compared to
            experimental results~\cite{zhang_performance_2018,
            berliner_effect_1986,felice_temperature_1977,kittel_intro_solid_2005}
            (solid horizontal lines), which have been corrected for zero-point
            vibrational energy using the HSE06 corrections from
            Ref.~\onlinecite{zhang_performance_2018}.  DFT results for the LDA
            and HSE06 functionals are shown for comparison (from
            Ref.~\onlinecite{zhang_performance_2018})
	}
	\label{fig:structplotli}
\end{figure}

Results are presented in Fig.~\ref{fig:structplotli}, where they are compared
to low temperature experimental results~\cite{zhang_performance_2018, berliner_effect_1986,felice_temperature_1977,kittel_intro_solid_2005}
that have been corrected for zero-point
vibrational effects based on HSE06 phonon calculations~\cite{zhang_performance_2018};
a table of all values is given in the Supplemental
Material~\cite{Note1}.  Consistent with our results on the UEG, we see relatively systematic
improvement with increasing sophistication of the theory. DCSD, ring-CCSDT,
CCSDT, and DCSDT are all improvements over CCSD and they achieve accuracies of
0.004--0.02~\AA, 0.1--0.2~GPa, and 4--6~m$E_h$ in the lattice constant, bulk
modulus, and cohesive energy, respectively.
It is hard to disentangle the
remaining discrepancies, which likely include some combination of pseudopotential,
basis set, and finite-size error, incomplete correlation, and experimental
uncertainty, including vibrational corrections.
We also compare to DFT results reported in Ref.~\cite{zhang_performance_2018} 
using the LDA~\cite{kohn_self-consistent_1965} and 
HSE06~\cite{heyd_hybrid_2003,heyd_erratum_2006,krukau_influence_2006} functionals.
While the LDA functional does not predict accurate structural properties
(despite its exactness for the UEG), the HSE06 functional performs very well.
Importantly, we see that the improved methods explored in this work clearly
outperform CCSD, bringing CC theory in line with the best performing DFT functionals.

\textit{Conclusion.}
Despite the apparent simplicity of simple metals, including the uniform
electron gas, achieving high accuracy for the electron correlation energy with
\textit{ab initio} wavefunction or diagrammatic methods is clearly a challenge.
We have shown that within the family of CC theories, the infinite-order
inclusion of connected triple excitations is essential, although semi-empirical
treatments of these effects are surprisingly effective.  We expect that the
methods explored here, which have been evaluated for their ability to predict
the properties of nearly uniform systems, will outperform DFT for more
heterogeneous systems, such as those arising in surface chemistry.  Before CC
methods are widely used in this context, their comparatively high
computational and storage costs must be addressed.  However, in the meantime,
they can be used to provide predictions of benchmark quality, especially in the
many situations where experimental values cannot be obtained to the required
precision.

\vspace{2em}

\textit{Acknowledgments.} 
We thank James Callahan and Xiao Wang for helpful discussions.  This work was
supported by the Columbia Center for Computational Electrochemistry and the
National Science Foundation under Grant No.~CHE-1848369.  We acknowledge
computing resources from Columbia University's Shared Research Computing
Facility project, which is supported by NIH Research Facility Improvement Grant
1G20RR030893-01, and associated funds from the New York State Empire State
Development, Division of Science Technology and Innovation (NYSTAR) Contract
C090171, both awarded April 15, 2010.
Data analysis and visualization were performed using 
NumPy~\cite{harris_array_2020},
SciPy~\cite{scipy_2020},
pandas~\cite{mckinney_data_2009}, 
Matplotlib~\cite{hunter_matplotlib_2007},
seaborn~\cite{Wseaborn2021}, and
JaxoDraw~\cite{binosi_jaxodraw_2004}. UEG calculations used
Julia~\cite{bezanson_julia_2017}, 
Fermi.jl~\cite{aroeira_fermijl_2022},
TensorOperations.jl~\cite{noauthor_tensoroperationsjl_nodate}, 
and Tullio.jl~\cite{noauthor_tulliojl_nodate}.

\end{document}


\title{Supplemental Material for: Highly accurate electronic structure of metallic solids from coupled-cluster theory with nonperturbative triple excitations}
\author{Verena A. Neufeld}
\author{Timothy C. Berkelbach}
\affiliation{Department of Chemistry, Columbia University, New York, New York 10027, USA}
\date{\today}  

\maketitle

\renewcommand\thesection{S\arabic{section}}
\renewcommand\theequation{S\arabic{equation}}
\renewcommand\thefigure{S\arabic{figure}}
\renewcommand\thetable{S\arabic{table}}

\section{Uniform electron gas calculations}
\label{sec:ueg}

In a plane-wave basis, the Hamiltonian of the uniform electron gas (UEG) in a cubic box of
volume $V=L^3$ is
\begin{align}
    H &= 
    \sum_{\vk\sigma} \epsilon_k a^\dagger_{\vk,\sigma} a_{\vk,\sigma}
    + \frac{1}{2}\sum_{\vk_1\sigma_1,\vk_2\sigma_2}\sum_{\vq\neq 0} v(q) 
        a^\dagger_{\vk_1+\vq,\sigma_1} a^\dagger_{\vk_2-\vq,\sigma_2} a_{\vk_2,\sigma_2} a_{\vk_1,\sigma_1} \\
    v(q) &= \begin{cases}
        4\pi e^2/(Vq^2) & q \neq 0 \\
        v_\mathrm{M} & q = 0
    \end{cases}
\end{align}
where $v_\mathrm{M} \propto L^{-1}$ is the Madelung constant of the cell, 
$\epsilon_k = \hbar^2k^2/(2m)$, $\vk = (2\pi/L)[n_x, n_y, n_z]+\vk_\mathrm{s}$ ($n_i$ are integers),
and $\vk_\mathrm{s}$ is a shift of the $k$-point mesh consistent with twisted boundary conditions.
Here we choose $\vk_\mathrm{s}$ to be the Baldereschi point~\cite{baldereschi_mean-value_1973} of the cubic Brillouin zone,
i.e., $\vk_\mathrm{s} = (2\pi/L)[1/4, 1/4, 1/4]$,
which we find to smooth the convergence to the
thermodynamic limit (TDL).  We study only closed-shell systems with ``magic
numbers'' of electrons $N$ increasing by roughly a factor of 2, i.e., $N =
14,34,70,\ldots,1404$. CCSD-based and CCSDT-based calculations are performed
using up to 1404 and 156 electrons, respectively. 
The box volume is based on the target density $n$, i.e., $V = N/n$.

We note that the orbitals to be used in constructing \textit{any}
reference determinant without translational symmetry breaking, such as spin- or
charge-density wave ordering, are just the plane-wave orbitals with the lowest
kinetic energy. In ``full'' CC theories, such as CCSD or CCSDT, the final
results depend only on this determinant of single-particle orbitals and
\textit{not} on their orbital energies. This point is important because the HF
orbital energies are known to be problematic for metals. However, removing some
terms from CC theory, as done in the (direct) RPA and ring-CCSDT, reintroduces
a dependence on the orbital energies.  The traditional RPA theory uses the kinetic
energies only; for consistency with standard CC
implementations, we use HF orbital energies in our ring-CCSDT calculations.

As a single-particle basis, we use $M$ plane-wave spin-orbitals, where the largest value of 
$M$ depends on $N$. Ultimately, we seek the combined complete basis set (CBS) limit
and thermodynamic limit (TDL), which we describe in Sec.~\ref{ssec:cbs} and \ref{ssec:tdl}.
But first, we demonstrate the convergence and divergence behaviors with $N$ of the CC
theories discussed in the main text.

\subsection{Demonstration of convergence or divergence behaviors}

\begin{figure}[t!]
	\includegraphics[width=3.5in]{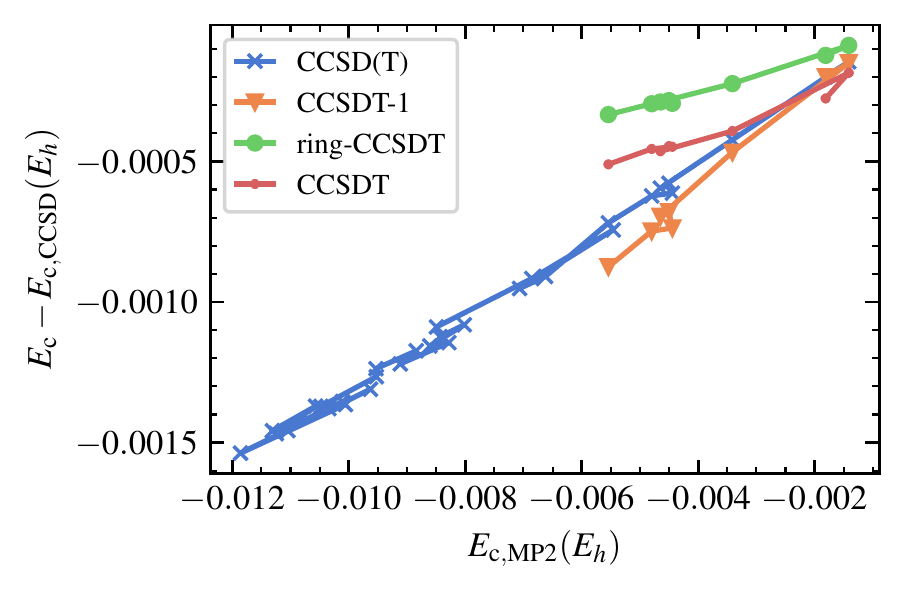}
	\caption{Thermodynamic limit convergence at $r_s = 4$ of the difference of the CCSD(T), CCSDT, CCSDT-1,
		and ring-CCSDT correlation energies to the CCSD correlation energy, plotted against the 
                MP2 correlation energy.
	        The ratio of the number of orbitals to number of
                electrons is $M/N\approx 1.5$, which is far from the complete basis set limit.
                Results are shown for systems ranging from $(N=34, M=52)$ to $(N=180, M=272)$ for
                CCSDT-1, ring-CCSDT, CCSDT, and up to $(N=628, M=960)$ for CCSD(T).}
	\label{fig:tdl_trip}
\end{figure}

In Fig.~\ref{fig:tdl_trip}, we follow Refs.~\onlinecite{shepherd_many-body_2013,shepherd_range-separated_2014,shepherd_coupled_2014} 
and plot the correlation energy of CCSD(T), CCSDT-1, ring-CCSDT, and CCSDT against
that of MP2 theory, with $M/N \approx 1.5$ (i.e.,
the results are not in the CBS, but are sufficient to demonstrate the convergence or divergence with $N$).
More specifically, we plot the difference with respect to the CCSD correlation energy, which is known
to converge with $N$.
For CCSD(T), this difference is exactly the (T) correction,
which we find to increase in linear proportion to the MP2 correlation energy, which is fully
consistent with our analytical conclusion that both diverge logarithmically.
We see that CCSDT-1 appears to diverge at least as fast as CCSD(T) and shows no indication
of convergence at the accessible values of $N$.
In contrast, ring-CCSDT and CCSDT show the onset of a plateau, providing strong
numerical support for their convergence in the TDL, which is consistent with our expectations. 

\subsection{Complete basis set limit estimation}
\label{ssec:cbs}

For each $N$, we estimate the $M\rightarrow \infty$ complete basis set (CBS) limit incrementally, 
using a composite correction based on the smaller value of $N$ at the same level of theory,
\begin{equation}
\label{eq:cbs}
E_{N}(\mathrm{CBS}) \approx E_{N}(M) + E_{N/2}(\mathrm{CBS}) - E_{N/2}(M/2),
\end{equation}
where $M$ is the largest value accessed for that value of $N$, ``$N/2$'' is
short-hand for the number of electrons that is \textit{roughly} a factor of 2
smaller than $N$, and $E_{N/2}(M/2)$ is found via interpolation of results with
varying $M$, since we are limited to ``magic numbers'' of basis functions.
Table~\ref{tab:maxsystemsforueg} shows, for each CC method,
the maximum number of spin-orbitals $M$ that we use for each number of electrons $N$.

\begin{table}[b]
	\centering
	\begin{tabular*}{0.7\textwidth}{@{\extracolsep{\fill}}lccccccc}
		\hline\hline
		& $N=14$ & $N=34$  & $N=70$ & $N=156$ & $N=332$ & $N=700$& $N=1404$ \\
		\hline
		CCSD & 47118& 23966  & 10332 & 5252 & 2488 & 2392 & 2392 \\
		DCSD & 52746 & 27154  & 19962 & 8646 & 4850 & 3996& 4140 \\
		CCSDT &1476--1538& 832  & 502 & 344 & - & -& - \\
		ring-CCSDT & 5720 & 1676  & 1104 & 302--410 & - & -& - \\
		DCSDT & 1476 & 796--832 & 464--502 & 272--344 & - & -& - \\
		\hline\hline
	\end{tabular*}
        \caption{Maximum number of spin-orbitals $M$ for given number of
        electrons $N$ for the CC methods.  The maximum number of spin-orbitals $M$ sometimes 
        depends on $r_s$, in which case a range is given.}
	\label{tab:maxsystemsforueg}
\end{table}

For CCSD-based methods, we use Eq.~(\ref{eq:cbs}) to
estimate the correlation energy $E_\mathrm{c}$. For the more expensive
CCSDT-based methods, we use Eq.~(\ref{eq:cbs}) to estimate
the \textit{difference} to the DCSD correlation energy, $\Delta
E_\mathrm{c}(X\text{SDT}-\text{DCSD}) \equiv E_\mathrm{c}(X\text{SDT}) -
E_\mathrm{c}(\text{DCSD})$. 
In Fig.~\ref{fig:ueg_cbs_tdl}(a), we show the
convergence of the incremental basis set corrections at $r_s=4$, which can be
seen to decrease in magnitude with increasing $N$. 
The final CBS energies and energy differences are then
used to estimate the TDL, as described next.

\subsection{Thermodynamic limit extrapolation}
\label{ssec:tdl}

\begin{figure}[h!]
	\includegraphics[width=6.5in]{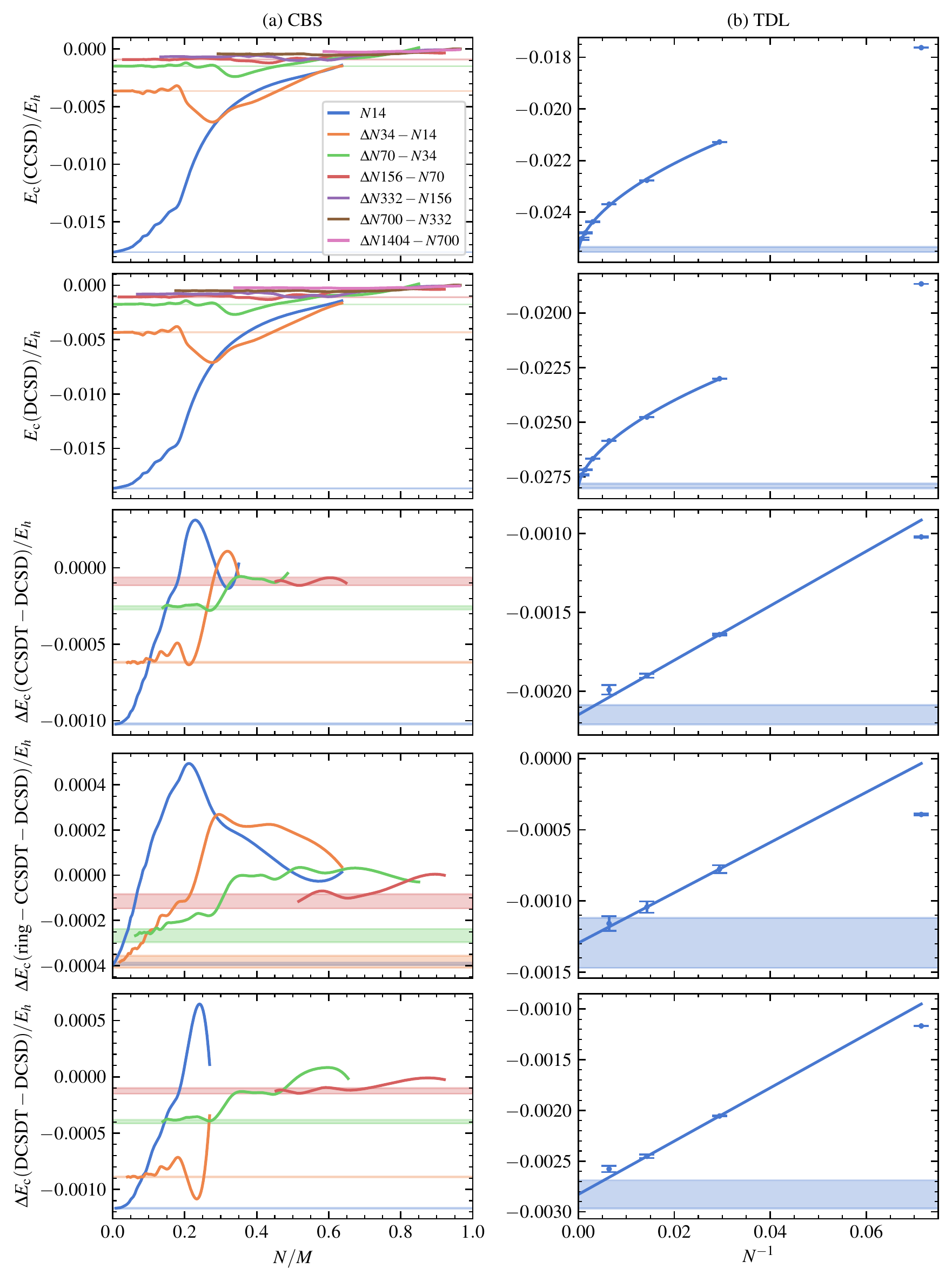}
	\caption{Convergence to (a) complete basis set (CBS) limit and (b) thermodynamic limit (TDL)
	for the uniform electron gas at $r_s = 4$. CCSD and DCSD correlation energies are shown, 
        as well as CCSDT, ring-CCSDT, and DCSDT energy differences to that of DCSD. 
        Estimated values with error bars are shown with shaded regions.}
	\label{fig:ueg_cbs_tdl}
\end{figure}

Given CBS estimates at each value of $N$, we estimate the TDL by extrapolation.
As justified in Sec.~\ref{sec:finite}, we extrapolate the CCSD-based correlation energies 
by fitting to the functional form
\begin{equation}
\label{eq:extrap}
E_\mathrm{c}(N) = E_\mathrm{c}(N\rightarrow \infty) + aN^{-2/3} + bN^{-1},
\end{equation}
where $E_\mathrm{c}(N)$ is the correlation energy per particle. In practice,
we fit to the six data points with the largest $N$.
For CCSDT-based methods, we extrapolate the CBS estimate of the \textit{difference} to DCSD,
as described in the previous section.
Because the $N^{-2/3}$ term in Eq.~(\ref{eq:extrap}) is attributed to the HF exchange energy,
which should cancel in the difference of two correlated energies, we extrapolate these
differences using only $N^{-1}$,
i.e., with $a=0$; specifically, a two-point extrapolation is performed with $N=34$ and $N=70$.
We then add this difference to the combined CBS+TDL value of the DCSD correlation energy
to obtain our final estimate of the CBS+TDL value of the CCSDT-based correlation energies.
Figure~\ref{fig:ueg_cbs_tdl}(b) shows the TDL convergence at $r_s = 4$ for the correlation
energy of CCSD and DCSD and for the correlation energy difference of various CCSDT-based methods.

\subsection{Final Results}
Combining the extrapolations described above yields our final correlation energy estimates in
the CBS and TDL limits, which are given in Tab.~\ref{tab:uegabsresults} and were presented
graphically in Fig.~1 of the main text.
\begin{table}[h]
\centering
\begin{tabular*}{0.48\textwidth}{@{\extracolsep{\fill}}lccccc}
\hline\hline
& \multicolumn{5}{c}{$E_\mathrm{c}/N$ (m$E_h$)} \\
& $r_s = 1$ & $r_s = 2$ & $r_s = 3$ & $r_s = 4$ & $r_s=5$  \\
\hline
CCSD & $-$56  & $-$39 & $-$31 & $-$25 & $-$22 \\
CCSDT & $-$59 & $-$44 & $-$35 & $-$30 & $-$26 \\
ring-CCSDT & $-$59 & $-$43  & $-$35  & $-$29 & $-$25  \\
SCS-CCSD & $-$68 & $-$48 & $-$38 & $-$31 & $-$27 \\
DCSD &$-$58 & $-$42 & $-$33 & $-$28& $-$24 \\
SCS-DCSD &$-$63 & $-$45& $-$36& $-$30 & $-$26 \\
DCSDT & $-$59 & $-$44 & $-$36& $-$31 & $-$27 \\
DMC~\cite{perdew_self-interaction_1981,ceperley_ground_1980} & $-$60& $-$45& $-$37& $-$32& $-$28 \\
\hline\hline
\end{tabular*}
\caption{Uniform electron gas correlation energy per electron for $r_s = 1-5$
from various coupled cluster methods
extrapolated to the complete basis set and thermodynamic limit.
(Fitted) diffusion Monte Carlo (DMC)
results~\cite{perdew_self-interaction_1981,ceperley_ground_1980} 
are shown for comparison.}
\label{tab:uegabsresults}
\end{table}

\section{Many-body finite-size errors in metals}
\label{sec:finite}

Here, we provide an analysis of asymptotic finite-size errors in metals, based
on the uniform electron gas (UEG).
The potential energy (per electron) of the UEG can be expressed via the Coulomb interaction
$v(q)=4\pi e^2/q^2$ and the structure factor $S(q)$~\cite{giuliani_quantum_2005},
\begin{equation}
U = 
\frac{1}{2} \int \frac{d^3q}{(2\pi)^3} v(q) \left[ S(q)-1 \right]
    = \frac{e^2}{\pi} \int_0^\infty dq
    \left[ S(q)-1 \right].
\end{equation}
The exact long wavelength behavior of the structure factor is
$S(q) = \hbar q^2 / 2m\omega_p$, where $\omega_p = \sqrt{4\pi ne^2/m}$ is the
plasmon frequency; we expect all correlated methods that include the physics of
the RPA capture this long wavelength behavior.
In numerical calculations of finite systems with
periodic boundary conditions, momentum transfers near $q=0$ are neglected, and
so we can estimate the finite-size error as 
\begin{equation}
\Delta U = \frac{e^2}{\pi} \int_0^{q_c} dq S(q)
\approx \frac{\hbar e^2}{6m\pi\omega_p} q_c^3
\end{equation}
where
\begin{equation}
q_c = \left(\frac{3}{4\pi}\right)^{1/3} \frac{2\pi}{L}
= \left(6\pi^2n\right)^{1/3} N^{-1/3} = 2^{1/3} \kF N^{-1/3}
\end{equation}
is the radius of a sphere with volume $(2\pi)^3/V$.
Therefore, the finite-size error of the potential energy is $O(N^{-1})$.
This analysis assumes that the structure factor itself has no finite-size error.

However, the \textit{correlation} energy is the difference between
the interacting and noninteracting (Hartree-Fock) energies,
\begin{equation}
E_\mathrm{c} = \frac{e^2}{\pi} \int_0^{\infty} dq
\left[S(q)-S^{(0)}(q)\right].
\end{equation}
In contrast to the behavior of interacting theories, the long wavelength
behavior of the noninteracting
structure factor is $S^{(0)}(q) = 3q/4k_\mathrm{F}$, such that the
finite-size error of the mean-field potential energy is $O(q_c^2) \sim O(N^{-2/3})$.
Therefore, in principle, the correlation energy inherits this finite-size
error and it vanishes as $N^{-2/3}$ in the large $N$ limit.
Specifically, using these leading-order terms in both the exact interacting
and noninteracting structure factors, we can estimate the finite-size error as
\begin{equation}
\label{eq:leading}
\Delta E_\mathrm{c}
= \frac{\hbar e^2}{6m\pi\omega_p} q_c^3 - \frac{3e^2}{8\pi k_\mathrm{F}}q_c^2,
\end{equation}
which is functionally equivalent to Eq.~(\ref{eq:extrap}).

As a tractable demonstration of this behavior, we consider
the textbook RPA correlation energy,
\begin{equation}
\label{eq:rpa_qc}
E_\mathrm{c}^{(\mathrm{RPA})}
= \lim_{q_\mathrm{c}\rightarrow 0} E_\mathrm{c}^{(\mathrm{RPA})}(q_\mathrm{c}),
\ \ \ \ \ E_\mathrm{c}^{(\mathrm{RPA})}(q_\mathrm{c}) 
= \frac{\hbar}{2\pi n} \int_{q_\mathrm{c}}^{\infty} \frac{4\pi q^2dq}{(2\pi)^3}
\int_0^\infty d\omega
\Big\{
    \ln\left[1-v(q)\chi^{(0)}(q,i\omega)\right] 
     + 
    v(q)\chi^{(0)}(q,i\omega)
\Big\},
\end{equation}
which can be evaluated by simple two-dimensional quadrature
and $\chi^{(0)}(q,i\omega)$ is the Lindhard function.
Again, we emphasize that the first term (containing the logarithm)
is the RPA potential energy, and second first term only serves to remove
the Hartree-Fock exchange energy, as needed for the definition of the
correlation energy.
At small $q_\mathrm{c}$, the second (exchange) term dominates and upon integration
yields $E_\mathrm{c}^{(\mathrm{RPA})}(q_\mathrm{c}) \sim q_\mathrm{c}^2 \sim N^{-2/3}$.

\begin{figure}[h]
	\includegraphics[width=7in]{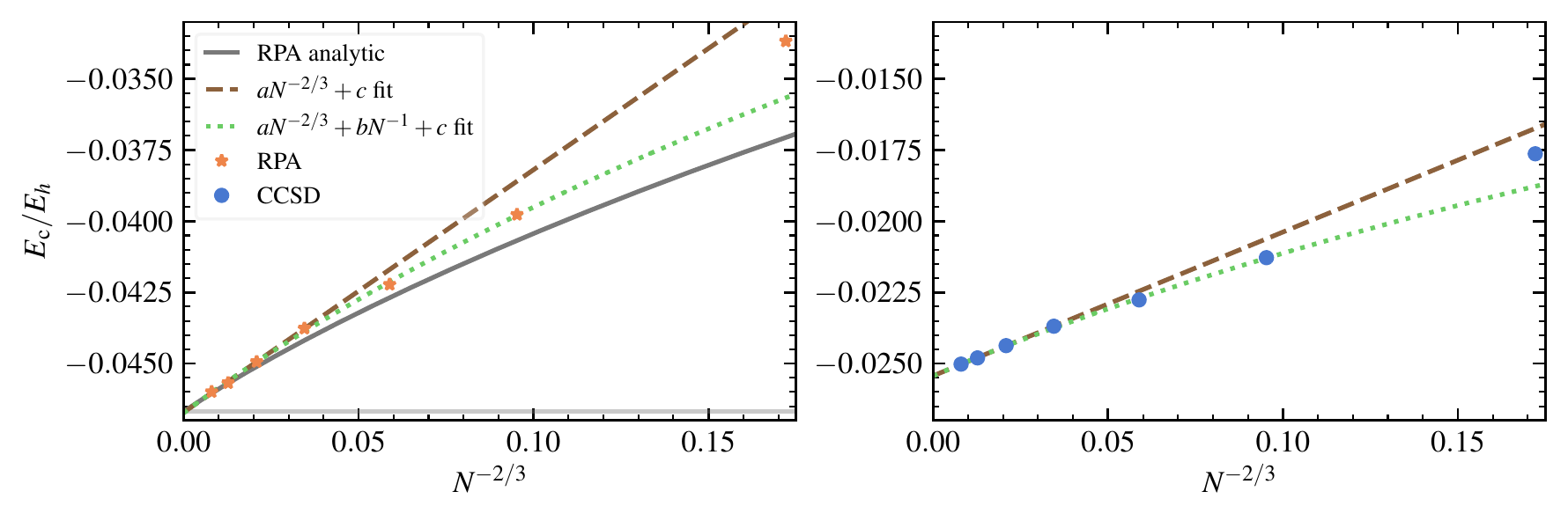}
	\caption{Thermodynamic limit convergence of the RPA (left) and CCSD (right)
		correlation energy at $r_s = 4$. 
	We show the numerical correlation energy from plane-wave based calculations (orange stars and blue circles).
        Two fits are tested on the results. For RPA, we also show the analytic correlation energy from Eq.~(\ref{eq:rpa_qc})
        (dark grey solid curve) and a horizontal line at the exact
        RPA correlation energy in the TDL.
        }
	\label{fig:tdl_drpa}
\end{figure}

This behavior is confirmed numerically, and the result is shown in Fig.~\ref{fig:tdl_drpa} (left)
for the UEG at $r_s=4$. We see that the analytic expression (evaluated by numerical integration)
converges to the TDL with the expected form. However, the asymptotic behavior (linear
in $N^{-2/3}$) is only observed for large system sizes with $N \gtrsim 100$. We show that the
same behavior persists for plane-wave based calculations performed in the manner of Sec.~\ref{sec:ueg}.
Fitting the correlation energy to Eq.~(\ref{eq:extrap}) using four data points (with $b=0$)
or with six data points (with $b\neq 0$) gives predictions that are in excellent agreement
with exact RPA result. We also show the CCSD TDL convergence in Fig.~\ref{fig:tdl_drpa} (right)
and similarly see that the fits with $b=0$ and $b\neq 0$ agree well with each other.
In principle, Eqs.~(\ref{eq:leading}) or (\ref{eq:rpa_qc}) could also be used effectively
as finite-size \textit{corrections}, although we do not do so here.

\section{Atomistic solid lithium calculations}

\subsection{Calculation details}
Except where otherwise stated, all calculation details, including basis sets and pseudopotentials,
are the same as in Ref.~\cite{neufeld_ground-state_2022}.
The CCSD data is taken from Ref.~\cite{neufeld_ground-state_2022}, but using $N^{-2/3}$ extrapolation
to the TDL, as discussed in Sec.~\ref{sec:finite}. As seen in Tab.~\ref{tab:libulk}, this difference
causes only small changes to our predicted properties.
Calculation details for bulk DCSD are similar to those from CCSD~\cite{neufeld_ground-state_2022},
except again for the TDL extrapolation. The single atom DCSD correlation energies were found using
two shells of ghost atoms in a molecular calculation.

As in our UEG study, the CC energies with triple excitations are estimated via their differences to DCSD energies
calculated with the same system size.
Specifically, calculations with triple excitations were performed on an 8-atom supercell
with $\Gamma$-point sampling (equivalent to sampling with four $k$ points on a two atom cubic unit cell)
and 1s core electrons frozen.
An additional calculation was performed
on a 16-atom cubic supercell at lattice parameter~3.5~\AA\ using frozen virtual natural orbitals. The difference
with respect to a calculation on the 8-atom supercell with the same frozen virtual treatment was used
as an additional finite-size correction.
At each basis set, the difference between the triple calculation and
DCSD at this system size was added to the DCSD energy at that basis set in the TDL.
Correlation energies were extrapolated to the CBS limit using a $X^{-3}$ 
form ($X=3,4$ for TZ,~QZ)~\cite{helgaker_basis-set_1997}.
After the TDL CBS results for each coupled cluster variant were estimated,
as in Ref.~\cite{neufeld_ground-state_2022}, the Birch-Murnaghan~\cite{birch_finite_1947,zhang_performance_2018}
equation-of-state was fit to extract the lattice parameter, bulk modulus, and cohesive energy.

\subsection{Final results}

Table~\ref{tab:libulk} shows our calculated structural and energetic properties of lithium, i.e.,
lattice parameter $a$, bulk modulus $B$, and cohesive energy $E_{\mathrm{coh}}$.
\begin{table}[h]
\centering
\begin{tabular*}{0.48\textwidth}{@{\extracolsep{\fill}}lccc}
\hline\hline
& $a$ (\AA) & $B$ (GPa) & $E_{\mathrm{coh}}$ (m$E_h$)  \\
\hline 
CCSD ($N^{-1}$)~\cite{neufeld_ground-state_2022}& 3.49 &12.8 & $-$51 \\
CCSD  & 3.48& 13.0&$-$52\\
DCSD & 3.47& 13.2&$-$55 \\ 
ring-CCSDT  & 3.46& 13.1 & $-$57\\
CCSDT  & 3.47& 13.3&$-$57 \\
DCSDT & 3.47 &13.2 & $-$57 \\
Experiment~\cite{zhang_performance_2018,berliner_effect_1986,felice_temperature_1977,kittel_intro_solid_2005}& 3.45& 13.3&$-$61 \\
\hline\hline
\end{tabular*}
\caption{Equilibrium lattice parameter $a$, bulk modulus $B$ and cohesive energy $E_{\mathrm{coh}}$ for bulk BCC Li. Shown
	are CCSD with a $N^{-1}$ TDL extrapolation from a previous study~\cite{neufeld_ground-state_2022}, CCSD (with a $N^{-2/3}$ extrapolation), DCSD,
	ring-CCSDT, CCSDT, and DCSDT as well as zero-point motion corrected experimental 
	results~\cite{zhang_performance_2018,berliner_effect_1986,felice_temperature_1977,kittel_intro_solid_2005}.}
\label{tab:libulk}
\end{table}